\documentclass{ifacconf}
\usepackage{epsfig,rotating,setspace,latexsym,amsmath,epsf,amssymb,amsfonts,bm,theorem,subfigure,epstopdf,graphicx}
\usepackage[normalem]{ulem}
\usepackage{dsfont}
\usepackage{color}
\usepackage{comment}
\newenvironment{Proof}[1]{\medskip\par\noindent{\bf Proof:\,}\,#1}{{\mbox{\,$\blacksquare$}\par}}

\DeclareMathOperator*{\argmin}{arg\,min}
\usepackage{natbib}        
\begin{document}
\begin{frontmatter}

\title{Remote Estimation Games with Random Walk Processes: Stackelberg Equilibrium\thanksref{footnoteinfo}} 

\thanks[footnoteinfo]{Research of the authors was supported in part by the ARO MURI Grant AG285 and in part by ARL-DCIST Grant AK868.}

\author[First]{Atahan Dokme} 
\author[Second]{Raj Kiriti Velicheti} 
\author[Third]{Melih Bastopcu}
\author[Second]{Tamer Ba\c{s}ar}
\address[First]{Electrical and Electronics Engineering Department, Bogazici University, 
   Bebek, Istanbul, 34342, Türkiye (e-mail: adokme2@illinois.edu).}
\address[Third]{Department of Electrical and Electronics Engineering, Bilkent University, 
   Ankara, 06800, Türkiye (e-mail: bastopcu@bilkent.edu.tr).}

\address[Second]{The Coordinated Science Laboratory, University of Illinois Urbana-Champaign, 
   Urbana, IL, 61801, USA (e-mail: \{rkv4,basar1\}@illinois.edu)}

\begin{abstract}                
Remote estimation is a crucial element of real time monitoring of a stochastic process. While most of the existing works have concentrated on obtaining optimal sampling strategies, motivated by malicious attacks on cyber-physical systems, we model sensing under surveillance as a game between an attacker and a defender. This introduces strategic elements to conventional remote estimation problems. Additionally, inspired by increasing detection capabilities, we model an element of information leakage for each player. Parameterizing the game in terms of uncertainty on each side, information leakage, and cost of sampling, we consider the Stackelberg Equilibrium (SE) concept where one of the players acts as the leader and the other one as the follower. By focusing our attention on stationary probabilistic sampling policies, we characterize the SE of this game and provide simulations to show the efficacy of our results. 
\vspace{-0.04cm}
\end{abstract}


\end{frontmatter}

\section{Introduction}\label{sec:intro}
\vspace{-0.2cm}
With the rapidly evolving landscape of smart devices, the design and analysis of cyber-physical systems is taking center stage. As systems get smarter, there is an increasing interest for real-time monitoring of time varying system states. However, such remote estimation paradigms are unfortunately vulnerable to cyber attacks due to distant access. Key to defenses in such paradigms are detection and estimation. Consider, for example, the case of an attacker with a private internal state trying to get information about a stochastic target system. While there is asymmetric information on both sides, remote access allows each player to sample the state of their opponent. However, as detection capabilities improve, each player must be careful not to leak personal information. This gives rise to an asymmetric information game.

Similar scenarios occur in many other situations such as autonomous driving, privacy-preserving decision making, and strategic warfare. Traditionally, real-time tracking would involve a source sending updates to a destination (or controller) about the state of the system. Although one would like the source to keep the destination as up-to-date as possible, practical considerations such as network capacity and transmission costs hinder such a possibility. To measure the freshness of information at the destination, the notion of age of information (AoI) has been introduced \cite{Kaul12a} which is defined as the time elapsed since the most recent update received at the destination has been generated at its source. 
The AoI literature has garnered significant attention and has been extensively studied over the past decade. Recent advances in the AoI literature have been comprehensively reviewed in a survey paper by \cite{Yates2021survey}. Of particular note are age-optimal scheduling policies under limited communication constraints. In multi-source systems, the Maximum-Age-First (MAF) policy has been shown to be optimal for systems that can transmit one update at a time \cite{Bedewy19, Kadota18a}. For a special-class of age-penalty functions, \cite{Zhong19} have demonstrated that the Maximum Weighted Age Reduction (MWAR) policy is optimal. The optimality of Whittle's index policy has also been examined in \cite{maatouk2020optimality}. Using Markov Decision Processes (MDPs),  asymptotically age-optimal policies have been derived in \cite{chen2021minimizing, hsu2019scheduling}.

All the works mentioned above consider age-based cost functions. However, in some applications, the receiver may need to track a specific dynamic process. For this purpose, remote estimation of Wiener and Ornstein-Uhlenbeck (O.U.) processes under limited sampling constraints has been studied, respectively, in \cite{nar2014sampling, Sun_Wiener}, and \cite{arafa2021sample}. The remote estimation of a single binary Markov process using an age of incorrect information metric \cite{maatouk2020age} has been examined in \cite{kam2020age}. Similarly, the remote estimation problem for multiple binary Markov processes with a Poisson-based updating method has been investigated in \cite{Bastopcuinfection2022}. Additionally, the remote estimation of a random walk process with a sampling cost has been explored in \cite{Yun18}, which is the most closely related work to ours.

Note that none of the situations above take surveillance into account, as is the case in the motivating example earlier in this section. Accounting for such strategic interactions while sampling, changes the strategy drastically. To start with, building upon \cite{VelichetiStonyBrook}, we introduce a formal model of remote estimation games parameterizing the asymmetry in information on both ends, a privacy level for both players, and cost of sampling. This game has an interesting feature in that because although each player has an incentive to sample the state of the other player, doing so leaks information about their own state, and hence causing a trade-off. Due to the natural commitment order in attacker and defender games, we compute the optimal sampling strategy to commit to for the defender. Finally, via extensive discussion and simulations, we identify situations when it is optimal for the defender to actively sample the attacker's state although it leaks information about his own state to the attacker.

This leads to dynamic games with asymmetric information on both sides. In their pioneering work, \cite{aumann1995repeated} has introduced a study on repeated games with asymmetric information on one side restricting attention to zero-sum games. Follow-up works have extended their results to asymmetry of information on both sides as in \cite{amitai1996repeated}. In \cite{sorin1983some} and \cite{horner2010markov}, the authors proved hardness results on few classes of general sum games and zero-sum Markov games. See \cite{mertens1990repeated} for a comprehensive survey of related works. In all of these works, the asymmetric information does not evolve with the game, and hence cannot naturally model remote sensing problems with stochastic state drift as is the case in our motivating example. Further, all these works deal with computation of Nash Equilibrium as opposed to Stackelberg Equilibrium (SE) as in our case. Our work is also related to asymmetric information pursuit-evasion games \cite{olsder1988use}. However, as opposed to those mentioned works, our work has asymmetric information at both the players. The novelty of analyzing this problem within a SE framework lies in the distinct roles assigned to the two players: one is designated as the leader, making decisions first, while the other follows, reacting to the leader's strategy. Furthermore, there exist various real-life interpretations for these roles. An attacker and a defender in a cyber-physical system, a buyer and a seller in a market, and a company and a competitor in a sector are such examples. However, restricting attention to stationary strategies helps us circumvent signaling aspects in the presence of asymmetric information. 

This paper is structured as follows: In Section~\ref{sec:System_model}, we introduce remote estimation games. In Section~\ref{sec: erroranalysis}, we find closed-form expressions for the objectives of each player for a class of
sampling policies. In Section~\ref{sec:main_results}, we characterize SE policies. We present extensive numerical results in Section~\ref{sec:num_results}. We provide a discussion on the SE policies and conclude our paper in Section~\ref{Sec:discussion}.

\vspace{-0.3cm}
\section{System Model}\label{sec:System_model}
\vspace{-0.3cm}
We consider the problem of timely tracking of two independent random walk processes related to players $P_1$ and $P_2$. The random walk of player $P_i$ ($i=1,2$) at time $t$ is denoted by $x_i(t)$, and evolves as
\begin{align*}
    \\[-3.4em]
\end{align*}
\begin{align*}
    x_i(t+1) = x_i(t) + w_i(t), 
\end{align*}
\begin{align*}
    \\[-3.4em]
\end{align*}
where $w_i(t)$ is given by 
\begin{align*}
    \\[-3.4em]
\end{align*}
\begin{align}
w_i(t) =  \begin{cases} 
      1, & \text{ with prob. $ \alpha_i$,} \\
      0, & \text{ with prob. $ 1-2\alpha_i$,} \\
      -1, & \text{ with prob. $ \alpha_i$,} 
   \end{cases}    
\end{align}
\begin{align*}
    \\[-3.2em]
\end{align*}
with $0<\alpha_i<0.5$ for $i=1,2$. Here, each player is interested in keeping track of the other player's state as \textit{timely} and \textit{accurately} as possible. With this goal, player $P_1$ (resp. player $P_2$) maintains an estimate of player $P_2$'s state, denoted by $\hat{x}_2(t)$ (resp. estimate of player $P_1$'s state, denoted by $\hat{x}_1(t)$). Then, the estimation error of player $P_i$ at time $t$ is given by $x_{e,i}(t) =x_{-i}(t)-\hat{x}_{-i}(t)$ where $-i$ refers to the player other than player $P_i$. In order to minimize the estimation error, player $P_i$ makes an observation of $x_{-i}(t)$ by taking action $u_{i}(t) = 1.$ Similarly, $u_{i}(t) = 0$ refers to not making any observation. We assume that whenever player $P_i$ takes a sample, it observes $x_{-i}(t)$ without any error or delay. However, it also reveals full information about its own state $x_i(t)$ to player $P_{-i}$, even if player $P_{-i}$ does not make any observations, that is, $u_{-i}(t) = 0$. With this information revelation mechanism, if we denote the message that player $P_i$ receives at time $t$ by $x_{m,i}(t)$, we have
\begin{align}\label{eqn:messaging_sch}
(x_{m,1}(t),&x_{m,2}(t)) = \nonumber\\ &\begin{cases} 
    \emptyset, & 
      \text{if $(u_1(t),u_2(t)) = (0,0)$,}\\
      (x_{2}(t),x_{1}(t)), &  \text{otherwise.}
   \end{cases}    
\end{align}
Each player wants to minimize his own long term estimation error while minimizing the cost of sampling and information revealed to the other player. The cost function of player $P_i$ is denoted by $J_i(\eta_1(t),\eta_{2}(t)) $ and is given by 
\begin{align*}
    \\[-3.5em]
\end{align*}
\begin{align}
J_1(\eta_1(t),\eta_{2}(t)) = &  \lim_{T\rightarrow\infty} \frac{1}{T} \sum_{t=1}^T \mathbb{E}[(x_2(t) -\hat{x}_2(t))^2] \nonumber\\[-0.2em]&
-\alpha \mathbb{E}[(x_{1}(t) -\hat{x}_{1}(t))^2]+c_1u_1(t),\label{eqn:obj_p1} \\[-0.2em]
J_2(\eta_1(t),\eta_{2}(t)) =&  \lim_{T\rightarrow\infty} \frac{1}{T} \sum_{t=1}^T \mathbb{E}[(x_{1}(t) -\hat{x}_{1}(t))^2]\nonumber\\[-0.2em]&- \alpha \mathbb{E}[(x_2(t) -\hat{x}_2(t))^2]]+c_2u_2(t), \label{eqn:obj_p2}
\end{align}
\begin{align*}
    \\[-3.2em]
\end{align*}
where $\eta_i(t)$ denotes the policy of player $P_i$ which maps from information available to player ($\mathcal{I}_i(t)$) to a sampling decision $u_i(t)$, $\alpha$ denotes the importance weight related to information revelation of players to their opponents, $c_1>0$ and $c_2>0$ denote the cost of taking a measurement for players $P_1$ and $P_2$, respectively. With the information acquisition mechanism in (\ref{eqn:messaging_sch}), the information structure of player $P_i$ is given by $\mathcal{I}_i(t)=\{x_i(t), x_{m,1}([t]),x_{m,2}([t]),u_{1}([t]), u_{2}([t]) \}$.\footnote{We let $[t] = \{1,\dots, t\}$ and $v([t]) = \{v(1),\dots, v(t)\}$ for any general state $v(t)$ parameterized by $t$.} We assume that the parameters of the game, $\{\alpha_1, \alpha_2, c_1, c_2, \alpha\}$, are known to both players.

Given the information structure as above, player $P_i$ forms the estimate $\hat{x}_{-i}(t)$. Due to the zero mean Markov evolution of the random walk process, the best estimator for each player is the most recent information about the state of the opponent, that is, $\hat{x}_{-i}(t) =x_{-i}(U_i(t))$ where $U_i(t) =\max\{t'| x_{-i}(t')\in \mathcal{I}_i(t)\}$. We use the expected MMSE error to characterize the accuracy of the estimation error which is given by $\mathbb{E}[x_{e,i}^2(t)]  =\mathbb{E}[(x_{-i}(t) -\hat{x}_{-i}(t))^2]$. In what follows, restricting attention to a class of stationary, probabilistic sampling policies, we obtain a simplified expression for the cost function of each player, then compute the best response of the follower for a given sampling strategy of the leader and then compute SE for the leader given the follower's best response.
\vspace{-0.35cm}
\section{Average Estimation Error Analysis}\label{sec: erroranalysis}
\vspace{-0.35cm}
As mentioned above, we concentrate on stationary probabilistic sampling policies where player $P_i$ makes a measurement with probability $0\!\leq\! p_i\!\leq\! 1$ at each time, independent from the current estimation error at the players.\footnote{Here, we would like to note that with the information structure $I_i(t)$, the players can come up with more sophisticated sampling policies which take the current estimation error into account, and thus the sampling policies can be in general in the form of $(p_1(t),p_2(t))$. However, in this work, we focus our attention on the naive probabilistic constant sampling policies and look for equilibrium strategies within these policies which would be easy to implement.}
\begin{figure}[t]
\centerline{\includegraphics[width=0.78\columnwidth]{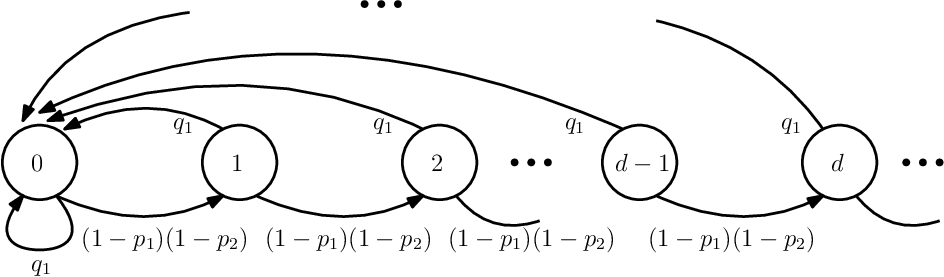}}
\vspace{-0.33cm}
	\caption{The Markov chain where the states represent the timeliness of the information at player $P_1$ with a given probabilistic sampling policy. Here, $q_1$ is equal to $q_1 = 1-(1-p_1)(1-p_2)$.}
	\label{Fig:MC}
 \vspace{-0.22cm}
\end{figure}
As the best estimator $\hat{x}_{-i}(t)$ is based on the most recent information available to at player $P_i$, we first characterize the information timeliness at the players by using the AoI metric. We denote the age of player $P_i$ at time $t$ as $\Delta_i(t) = t- U_i(t)$ where $U_i(t)$ is the time-stamp of the most recent information at player $P_i$. In this section, we first focus our attention on finding the stationary distribution of the AoI at player $P_1$. With the proposed probabilistic sampling policy, it is easy to verify that $\Delta_1(t)$ constructs a Markov chain with states $\mathcal{S} = \{0,1,2,\dots,\}$ as illustrated in Fig.~\ref{Fig:MC}. For a given age $\Delta_1(t-1)$ at time $t-1$, $\Delta_1(t)$ evolves as follows:     
\begin{align}
   \Delta_1(t) \!=\! \begin{cases} 
      0 & \text{w. p. $1- (1-p_1)(1-p_2)$,}\\
      \Delta_1(t-1) + 1 & \text{w. p. $(1-p_1)(1-p_2)$,}\!\!\!\!\!\!\!\!\!\!\!\! 
   \end{cases}
\end{align}
since age only increases when both the players do not sample, which happens with probability $(1-p_1)(1-p_2)$, and would go to zero otherwise.

We denote the steady-state distribution of AoI of player $P_1$ by $\pi_{1,k}$ for $k=0,1,2,\dots.$ If we write the state-balance equations and use $\sum_{k=0}^{\infty}\pi_{1,k} = 1$, we obtain the steady state distribution as \cite{bertsekas2008introduction}: 
\begin{align}
 \pi_{1,k} = (1-(1-p_1)(1-p_2))((1-p_1)(1-p_2))^k.
\end{align}
Next, we want to characterize the long term average estimation error $ \mathbb{E}[x_{e,1}^2] \!=\!\lim_{T\rightarrow\infty} \frac{1}{T} \sum_{t=1}^T\! x_e(t)^2$. For that, we first find $ \mathbb{E}[x_{e,1}^2|\Delta_1(t) = \Delta]$ which is the second moment of the error conditioned on $\Delta_1(t) = \Delta$, and is given by 
\begin{align*}\\[-2em]
   \mathbb{E}[x_{e,1}^2|\Delta_1(t) = \Delta]=\lim_{T\rightarrow\infty} \frac{1}{T_\Delta(T)} \sum_{t=1}^T x_{e,1}(t)^2\mathds{1}\{\Delta_1(t) = \Delta\}, \\[-1.8em] 
\end{align*}
where $T_\Delta(T) = \sum_{t=1}^T \mathds{1}\{\Delta_1(t) = \Delta\}$. Player $P_1$ has the accurate information when $\Delta_1(t) = 0$. When $\Delta_1(t) =\Delta > 0$, player $P_1$ had the accurate information about player $P_2$'s state at $t-\Delta_1(t)$. After that, at each time, player $P_2$'s state can increase by 1 or decrease by 1 with probability $\alpha_2$, and stays the same with probability $1-2\alpha_2$. 

We define $m_1$, $m_{-1}$, and $m_0$ as the total number of forward movements, backward movements, and staying in the same position within the time frame of  $t-\Delta$ and $t$, respectively. Then, the random variables $(M_1,M_0,M_{-1})$ have a multinomial distribution with probabilities $(\alpha_2, 1-2\alpha_2, \alpha_2)$ with the p.m.f. given by 
\begin{align}\label{eqn:multinom}
 \mathbb{P}(M_1=m_1,&M_0=m_0,M_{-1}=m_{-1})=\nonumber\\ &\binom{\Delta}{m_1,m_0,m_{-1}} \alpha_2^{m_1}(1-2\alpha_2)^{m_0} \alpha_2 ^{m_{-1}},  
\end{align}
where $m_1+m_0+m_{-1}=\Delta$ and $m_{i}\geq 0$ for $i=-1,0,1$. The first and second moments of the multinomial distribution are given by $\mathbb{E}[M_1] =\mathbb{E}[M_{-1}] = \Delta \alpha_2$, $\mathbb{E}[M_1^2] =\mathbb{E}[M_{-1}^2] = \Delta \alpha_2(1-\alpha_2) + \Delta^2 \alpha_2^2$, and $\mathbb{E}[M_1 M_{-1}] =-\Delta \alpha_2^2+\Delta^2\alpha_2^2$. 
For a given $\Delta_1(t) = \Delta$, we have $x_{e,1}(t)=x_2(t)-x_2(t-\Delta) = M_1 -M_{-1}.$ Thus, $\mathbb{E}[x_{e,1}^2|\Delta_1 \!\!=\!\! \Delta]$ is given by 
\begin{align*}
    \mathbb{E}[x_{e,1}^2|\Delta_1 \!\!=\!\! \Delta]\!\! =\!\! \mathbb{E}[M_1^2] \!-\!2\mathbb{E}[M_1M_{-1}] \!+ \!\mathbb{E}[M_{-1}^2]\! = \!2\Delta\alpha_2.
\end{align*}
By using iterative expectations, i.e., $\mathbb{E}[x_{e,1}^2] = \mathbb{E}[\mathbb{E}[x_{e,1}^2|\Delta_1]]$
\begin{align}
\mathbb{E}[x_{e,1}^2] = &\frac{2\alpha_2 (1-p_1) (1-p_2)}{1- (1-p_1)(1-p_2)}.
\end{align}
By following similar steps, we can obtain the average error for player $P_2$. Hence, the closed-form expressions for the cost functions in (\ref{eqn:obj_p1}) and (\ref{eqn:obj_p2}) are given by:
\begin{align}
J_1(p_1,p_2) = &  \frac{2(\alpha_2-\alpha \alpha_1) (1-p_1) (1-p_2)}{1- (1-p_1)(1-p_2)}+c_1p_1,\label{eqn:obj_p1_cls} \\
J_2(p_1,p_2) =&  \frac{2(\alpha_1-\alpha \alpha_2) (1-p_1) (1-p_2)}{1- (1-p_1)(1-p_2)}+c_2p_2.\label{eqn:obj_p2_cls}
\end{align}
As we focus our attention on the probabilistic sampling policies, we replace $J_i(\eta_1(t),\eta_2(t))$ with $J_i(p_1,p_2)$. Here, the goal of each player is to minimize their own cost functions provided in (\ref{eqn:obj_p1_cls}) and (\ref{eqn:obj_p2_cls}). Due to non-aligned objective functions of the players and information revelation mechanism, we have a non-zero sum game between the players which will be formulated as a Stackelberg game between the players defined precisely in the next section.

\vspace{-0.2cm}

\section{Equilibrium Characterization}\label{sec:main_results}
\vspace{-0.2cm}
As motivated in Section~\ref{sec:intro}, in this work, our goal here is to characterize the SE of the game, formally defined as:
\begin{defn}\label{def: Stackeq}
    A sampling probability pair $(p_1^*,p_2^*)$ constitutes a SE with player $P_1$ constitutes a leader and player $P_2$ as follower, each having their own cost functions $J_1(p_1,p_2)$ and $J_2(p_1,p_2)$  given by (\ref{eqn:obj_p1_cls}) and (\ref{eqn:obj_p2_cls}), resp. if
    \begin{align}
        p_1^*&=\argmin_{p_1\in[0,1]} J_1(p_1,BR(p_1)),\\
        \text{where}\quad& BR(p_1)=\argmin_{p_2\in[0,1]} J_2(p_1,p_2).
    \end{align}
    \begin{align*}
        \\[-38pt]
    \end{align*}
\end{defn}
Note that, we inherently assumed in Definition~\ref{def: Stackeq} that $J_2(p_1,p_2)$ has a unique minimum for each $p_1\in [0,1]$, which is true in our game. We define $K_1 =\frac{2(\alpha_2-\alpha\alpha_1)}{c_1} $ and $K_2 =\frac{2(\alpha_1-\alpha\alpha_2)}{c_2} $ which will appear in the equilibrium characterization in the remainder of this section. Next, we consider 2 different cases, namely 
(a) $K_2\leq0$ and (b) $K_2>0$. We will see that each case will lead to structurally different SE solutions. First, we consider the case $K_2\leq0$ and provide the SE in the following theorem.  
\begin{thm}\label{theorem_2}
    When $K_2\leq0$,  the SE of the game is given by
    \begin{align} \label{thm2_SE}
        (p_1^*,p_2^*) = \left(\min\left\{\sqrt{\max\{K_1,0\}},1\right\},0\right).
    \end{align}
\end{thm}
\vspace{-0.3cm}
\begin{Proof}
We first compute
    \begin{align*}\\[-1.8em]
    \frac{\partial J_2(p_1,p_2)}{\partial p_2}=c_2\left(-\frac{K_2(1-p_{1})}{(1-(1-p_2)(1-p_{1}))^2}+1\right).
    \end{align*}
Thus, when $K_2\leq0$, the objective function is an increasing function of $p_2$ as $\partial J_2(p_1,p_2)/\partial p_2>0$. In this case, the best response of player $P_2$ for any strategy of player $P_1$ is $p_2^*\!= \!0$.

Next, we turn our attention to player $P_1$'s optimization problem given that the follower's best response is always equal to $p_2^* =0$. For that, we substitute $p_2^* =0$ in $J_1(p_1,p_2)$ given in (\ref{eqn:obj_p1_cls}) and obtain 
    \begin{align}\label{eqn:cost_thm2}
        J_1(p_1,0) &= c_1\left(K_1\left( p_1^{-1}-1\right)+p_1\right).
    \end{align}
Then,
    \begin{align}\label{eqn:cost_thm2_der}
        \frac{\partial J_1(p_1, 0)}{\partial p_1}&=c_1 \left(-K_1 p_1^{-2}+1\right).
    \end{align}
    One can clearly observe that the optimal selection of $p_1$ depends on the sign of $K_1$. When $K_1\leq0$, we have $\partial J_1(p_1, 0)/\partial p_1>0$, and thus the optimal selection of $p_1$ is $p_1^* \!=\!0$. When $K_1\!>\!0$, we have $\partial^2 J_1(p_1, 0)/\partial p_1^2\!>\!0$, i.e., the cost function is convex; as a result, there exists a global minimum satisfying the first order necessary and sufficient conditions. Thus, by using $\partial J_1(p_1, 0)/\partial p_1=0$ and the fact that $p_1\!\in\![0,1]$, we have $p_1^* \!=\! \min\{\sqrt{K_1},1\}.$ Combining with the case $K_1\!\leq\!0$, we obtain the SE $(p_1^*,p_2^*)$ given in (\ref{thm2_SE}).\! \!\!\!
\end{Proof}
Therefore, when $K_2\leq0$, player $P_2$ always chooses $p_2^* = 0$, meaning that although player $P_1$ is the leader, it can not drive player $P_2$ away from not sampling. This could be due to the fact that $K_2\leq0$ implies $\alpha_1-\alpha \alpha_2\leq0$ which means that player $P_2$ places more importance on not revealing information than minimizing its estimation error. 
\begin{cor}\label{cor_3}
    If $K_1\leq0$ and $K_2\leq0$, the SE of the game is given by $(p_1^*, p_2^*) = (0,0)$.
    \end{cor}
Corollary~\ref{cor_3} simply follows from Theorem~\ref{theorem_2}. When $K_1\leq0$ and $K_2\leq0$, both players prioritize not revealing any information to the other player. For that, they prefer not to take any sample. When $K_1<0$ and $K_2<0$, although their own estimation error diverges, the cost function $J_i(p_1,p_2)$ for $i=1,2$ tends to negative infinity. When $K_1=0$ and $K_2=0$, the cost function $J_i(p_1,p_2)$ for $i=1,2$ equals to zero. Next, we consider the case $K_2>0$. \\ 
When $K_2 > 0$, 
the best response of player $P_2$ for a given player $P_1$'s policy of $p_1$ can be obtained by solving the following first-order condition: 
    \begin{align}
        \frac{\partial J_2(p_1,p_2)}{\partial p_2}=c_2\left(-\frac{K_2(1-p_{1})}{(1\!-\!(1\!-\!p_2)(1\!-\!p_{1}))^2}\!+\!1 \right) = 0,
    \end{align}
which yields
    \begin{align}
        p_2^*(p_1) = BR(p_1)&=1-\frac{1}{1-p_1}\left(1\!-\!\sqrt{K_2(1-p_1)}\right).
    \end{align}
The expression above is the best response function of Player 2 ($p_2^*$) for a given $p_1$. However, we need to impose the feasibility constraint $ 0\leq p_2^*\leq 1$, which leads to
    \begin{align}\label{eqn:ineq_for_p_2}
        0 \leq 1- (1-p_1)^{-1}\left[1-\sqrt{K_2(1-p_1)}\right] \leq 1.
    \end{align}
Solving the inequality above provides us a lower bound $p_1^L$ and an upper bound $p_1^U$. The lower bound $p_1^L$ can be obtained by solving $p_2^* =1$ and combining with $p_1^L\geq 0$, which leads to
        \begin{align}\label{eqn_p_1^L}
            p_1^L(K_2) = \max\left\{0, 1-K_2^{-1}\right\}.
        \end{align}
The upper bound $p_1^U$ can be similarly obtained by solving $p_2^* =0$ and using the fact that $p_1^U\leq 1$, which yields 
        \begin{align}
            p_1^U(K_2) = \frac{-K_2 + \sqrt{K_2^2+4K_2}}{2}.
        \end{align}
Here, we note that $p_1^U(K_2)$ is always less than 1. 
The critical boundary points $p_1^L$  and $p_1^U$ above constitute the best response function $p_2^* = BR(p_1)$ as follows: 
    \begin{align}\label{eqn:best_response_b2}
    \!\!p_2^*\!\!  = \!\!   \begin{cases} 
      1, & \text{if } 0\leq p_1<p_1^L, \\
      1\! -\! \frac{1}{1\!-\!p_1}\left(1\! \!-\!\! \sqrt{K_2(1\!-\!p_1)}\right), & \text{if } p_1^L\! \leq p_1 \!< p_1^U, \\
      0, & \text{if } p_1^U \leq p_1 \leq 1.
   \end{cases}    
   \end{align}
Due to (\ref{eqn_p_1^L}), if $p_1^L$ is equal to 0, then the first region, i.e., $0\leq p_1<p_1^L$, given in (\ref{eqn:best_response_b2}) disappears. Thus, when $p_1^L = 0$, we should only consider the second and third regions provided in (\ref{eqn:best_response_b2}). In the next three lemmas, we consider each region provided in (\ref{eqn:best_response_b2}), and find the optimal $p_1$ selection within each region. 
\begin{lem}\label{Lemma_3}
If $p_1^*\in[0,p_1^L]$, due to (\ref{eqn:best_response_b2}), we have $p_2^* = 1$. Then, the objective function of player $P_1$ in (\ref{eqn:obj_p1_cls}) becomes 
\begin{align}\label{eqn:obj_ply_1_case_1}
        J_1(p_1,1) =c_1p_1,
\end{align}
which is minimized when $p_1^*=0$.
\end{lem}
\vspace{-0.3cm}
\begin{Proof}
    When $p_1^*\in[0,p_1^L]$, it can be directly observed from (\ref{eqn:best_response_b2}) that $p_2^*$=1. As a result, the objective function of player $P_1$ is given by $J_1(p_1,1) =c_1p_1$, a linearly increasing function of $p_1$. Thus, $p_1*=0$ minimizes $J_1(p_1,1)$.  
\end{Proof}
As a result, if player $P_1$'s policy falls into the first region in (\ref{eqn:best_response_b2}), that is $p_1^*\in[0,p_1^L]$, the optimum choices of the players will be $(p_1^*,p_2^*)=(0,1)$. Next, we consider the third region in (\ref{eqn:best_response_b2}), that is $p_1^*\in[p_1^U,1]$, and find the optimal policies of the players. 

    \begin{lem}\label{Lemma_4}
     If $p_1^*\in[p_1^U,1]$, due to (\ref{eqn:best_response_b2}), we have $p_2^*= 0$. Then,  the objective function of player $P_1$ in (\ref{eqn:obj_p1_cls}) becomes
         \begin{align}\label{eqn:obj_ply_1_case_3}
        J_1(p_1,0) =c_1\left(K_1\left(p_1^{-1}-1\right)+p_1\right).
         \end{align}
To minimize $J_1(p_1,0)$, the optimal selection of $ p_1$ is (a) $p_1^*=p_1^U$ when $K_1\leq0$, and (b) $p_1^* = \min(\max(\sqrt{K_1},p_1^U),1)$ when $K_1>0$.
    \end{lem}
    \vspace{-0.3cm}
\begin{Proof}
   If $p_1^*\in[p_1^U,1]$, as provided in (\ref{eqn:best_response_b2}), the best response of player $P_2$ is $p_2^*= 0$. After substituting $p_2^*= 0$, we obtain $J_1(p_1,0)$ in (\ref{eqn:obj_ply_1_case_3}). Then, the derivative of $J_1(p_1,0)$ with respect to $p_1$ is the same as in (\ref{eqn:cost_thm2_der}) provided in Theorem~\ref{theorem_2}.  
The equilibrium structure is similar, but with one difference: the feasibility region of $p_1$ is now restricted to $p_1^*\in[p_1^U,1]$. When $K_1\leq0$, $\partial J_1 (p_1, 0)/\partial p_1>0$, and thus it is optimal to choose $p_1^*=p_1^U$. When $K_1>0$, since $J_1 (p_1, 0)$ is a convex function of $p_1$, the solution to $\partial J_1 (p_1, 0)/\partial p_1 =0$ and 
 considering $p_1^*\in[p_1^U,1]$ gives the optimal selection of $p_1$,
 which is $p_1^* = \min\{\max\{\sqrt{K_1},p_1^U\},1\}$, thus completing the proof.  
\end{Proof}
Thus, if player $P_1$'s policy falls into the last region in (\ref{eqn:best_response_b2}), that is $p_1^*\!\!\in\![p_1^U,1]$, the optimum choices of the
players 
will be $(p_1^*,p_2^*)\!=\!(p_1^U,0)$ if $K_1\!\leq\!0$, and $(p_1^*,p_2^*)=(\min\{\max\{\sqrt{K_1},p_1^U\},1\},0)$ if $K_1\!>\!0$.

In the next lemma, we consider the last remaining region, $p_1^*\in[p_1^L,p_1^U]$, and provide the optimal selection of $p_1^*$.  

\begin{lem}\label{Lemma_5}
If $p_1^*\in[p_1^L,p_1^U]$, after substituting the best response of player $P_2$ provided in (\ref{eqn:best_response_b2}), the cost function of player $P_1$ becomes 
    \begin{align}\label{eqn:obj_ply_1_case_2} 
    J_1(p_1,BR(p_1)) =c_1\left( \frac{K_1}{\sqrt{K_2}}{(1-p_1)}^{-\frac{1}{2}} - K_1 + p_1\right).\!\!
    \end{align}
In this case, the optimal selection of $(p_1,p_2)$ is given by
\begin{align}\label{eqn:opt_p1_p2_case_b}
   \!\!\!(p_1^*, p_2^*)\!\! =\!\!\begin{cases} 
      \!\argmin\{\!J_1(p_1^L,BR(\!p_1^L\!)), \!J_1(p_1^U,BR(p_1^U\!))\}, \\\hspace{4.65cm} \text{if $K_1<0$,}\\
      (p_1^L,BR(p_1^L))  \hspace{2.65cm} \text{if $K_1\geq0$,}
   \end{cases} \!\!\!\!
\end{align}
with $BR(p_1^L) = \min\{\sqrt{K_2},1\}$ and $ BR(p_1^U) = 0$ as in (\ref{eqn:best_response_b2}).
\end{lem}
\vspace{-0.3cm}
\begin{Proof}
Due to (\ref{eqn:best_response_b2}), the best response of player $P_2$ when $p_1^*\in[p_1^L,p_1^U]$ is given by $p_2^* = 1\! -\! \frac{1}{1-p_1}\left(1\! -\! \sqrt{K_2(1\!-\!p_1)}\right).$ 
After substituting $p_2^*$ into the cost function of player $P_1$, we obtain $J_1(p_1,BR(p_1))$ provided in (\ref{eqn:obj_ply_1_case_2}).
The first derivative of $J_1(p_1,BR(p_1))$ in (\ref{eqn:obj_ply_1_case_2}) with respect to $p_1$ is 
\begin{align}\label{eqn:Lem_5_first_der}
\frac{\partial J_1(p_1,BR(p_1))}{\partial p_1}=c_1\left(\frac{K_1}{2\sqrt{K_2}}{(1-p_1)}^{-\frac{3}{2}}+1\right),
\end{align}
and the second derivative is also given by  
\begin{align*}
\frac{\partial^2 J(p_1,BR(p_1))}{\partial p_1^2}=\frac{3}{4}\frac{c_1K_1}{\sqrt{K_2}}{(1-p_1)}^{-\frac{5}{2}}.\\[-18pt]
\end{align*}
Depending on the sign of $K_1$, the characteristic of the cost function $J(p_1,BR(p_1))$ changes. If $K_1<0$,  $J(p_1,BR(p_1))$ is a concave function of $p_1$. Thus, the minimum value of the cost function $J(p_1,BR(p_1))$ is obtained at one of the endpoints: either $p_1^L$ or $p_1^U$. Then, the optimal selection of players is $(p_1^*, p_2^*) =\argmin\{J_1(p_1^L,BR(p_1^L), J_1(p_1^U,BR(p_1^U)\}$ constituting the first part of the solution provided in (\ref{eqn:opt_p1_p2_case_b}). If $K_1\geq0$, the objective function $J(p_1,BR(p_1))$ in (\ref{eqn:obj_ply_1_case_2}) is an increasing function of $p_1$. Thus, in this case, the minimum is obtained when $(p_1^*,p_2^*) = (p_1^L,BR(p_1^L))$.  Combining these two cases in (\ref{eqn:opt_p1_p2_case_b}) concludes the proof.     
\end{Proof}

Eventually, $K_2>0$ case can be viewed as a piece-wise objective function for player $P_1$ in which $p_1^L$ and $p_1^U$ determine the endpoints. By using Lemmas~\ref{Lemma_3}, \ref{Lemma_4}, and \ref{Lemma_5}, we can write $J_1(p_1, BR(p_1))$ as follows:   
    \begin{align}\label{eqn:cost_player_1_K_2>0}
    \!\!\!J_1(p_1, BR(p_1)) \!\!=\!\! \begin{cases} 
      c_1p_1,  \hspace{1.8cm} \text{if } 0\leq p_1<p_1^L,\!\! \\
      c_1\!\left(\frac{K_1}{\sqrt{K_2}}(1\!-\!p_1)^{-\frac{1}{2}} \!-\! K_1 \!+\! p_1\right)\!, \\ \hspace{2.5cm} \text{if } p_1^L \leq  p_1 \leq  p_1^U\!,\!\! \\
      c_1\left(K_1\left(\frac{1-p_1}{p_1}\right)+p_1\right), \\ \hspace{2.6cm} \text{if } p_1^U < p_1 \leq 1.\!\!
   \end{cases}    
   \end{align}
In the next theorem, we state the SE when $K_2>0$.

\begin{thm}\label{thm2}
If $K_2\geq 0$, depending on the sign of $K_1$, the SE can be obtained as follows: 
     \begin{align}\label{eqn:opt_p1_p2_case_b_overall}
   (p_1^*, p_2^*) =\begin{cases} 
      \argmin_{(p_1, p_2)\in \mathcal{A}}J_1(p_1, p_2), & \text{if $K_1<0$, }\\
      \argmin_{(p_1, p_2)\in \mathcal{B}}J_1(p_1, p_2), & \text{if $K_1\geq0$, }      
   \end{cases} 
   \end{align}
   where $\mathcal{A}=\{(0,BR(0)),(p_1^L,BR(p_1^L)), (p_1^U,BR(p_1^U))\}$ and  $\mathcal{B}\!=\! \{(0,BR(0)),(p_1^L,BR(p_1^L)),(\min(\max(\! \sqrt{K_1},p_1^U\! ),1), 0)  \} $.
  
\end{thm}
The proof directly follows from combining the analyses provided in Lemmas~\ref{Lemma_3}, \ref{Lemma_4}, and \ref{Lemma_5} and determining the global minimum of the objective function $J_1(p_1, BR(p_1))$ in (\ref{eqn:cost_player_1_K_2>0}) when $K_2>0$. Thus, when we combine all these cases, we see from Theorem~\ref{thm2} that when $K_2>0$, there are 3 potential $(p_1,p_2)$ candidates given in set $\mathcal{A}$ when $K_1<0$ and in set $\mathcal{B}$ when $K_1\geq 0$ to be the SE. Thus, we choose the one that gives the minimum cost for player $P_1$ among these 3 candidates. In the following corollary, we specify the SE when $K_1>0$ and $K_2>1$.

    \begin{cor}\label{Cor_8}
    If $K_1>0$ and $K_2>1$, the SE of the game is given by $(p_1^*, p_2^*) = (0,1)$.
    \end{cor}
    \vspace{-0.3cm}
    \begin{Proof}
    When $K_2>0$, $J_1(p_1, BR(p_1))$  consists of at most 3 partial functions provided in (\ref{eqn:cost_player_1_K_2>0}). When $K_2>1$, all these 3 distinct regions exist. From Lemma~\ref{Lemma_3}, $(p_1,p_2) = (0,1)$ minimizes the first region, i.e., $p_1\in[0,p_1^L]$, with the minimum value $J_1(0,1) = 0$.  Next, it is easy to verify that the objective function $J_1(p_1, BR(p_1))$ in (\ref{eqn:cost_player_1_K_2>0}) is piece-wise continuous at point $p_1^L = 1-\frac{1}{K_2}$ when $K_2>1$. Furthermore, in the second region, that is, $p_1\in[p_1^L, p_1^U]$, when $K_1>0$, the objective function is an increasing function as shown in (\ref{eqn:Lem_5_first_der}). Thus, in the first two regions, i.e., $0\leq p_1 \leq p_1^U$, $J_1(p_1, BR(p_1))$ is increasing and piece-wise continuous at $p_1^L$. Therefore, $(p_1,p_2) = (0,1)$ is the minimizer of $J_1(p_1, BR(p_1))$ in $0\leq p_1 \leq p_1^U$. 
    Finally, for the third region, i.e., $p_1\in [p_1^U,1]$, we can observe from (\ref{eqn:cost_player_1_K_2>0}) that $J_1(p_1, BR(p_1))>0$ when $K_1>0$. Hence, $J_1(p_1, BR(p_1))$ is minimized at $(p_1^*,p_2^*)=(0,1)$, which is the SE when $K_1>0$ and $K_2>1$. 
    \end{Proof}

\vspace{-0.2cm}
\section{Numerical Results}\label{sec:num_results}
\vspace{-0.2cm}
\begin{figure}[t]
	\begin{center}
	    \subfigure[]{%
			\includegraphics[scale=0.17]{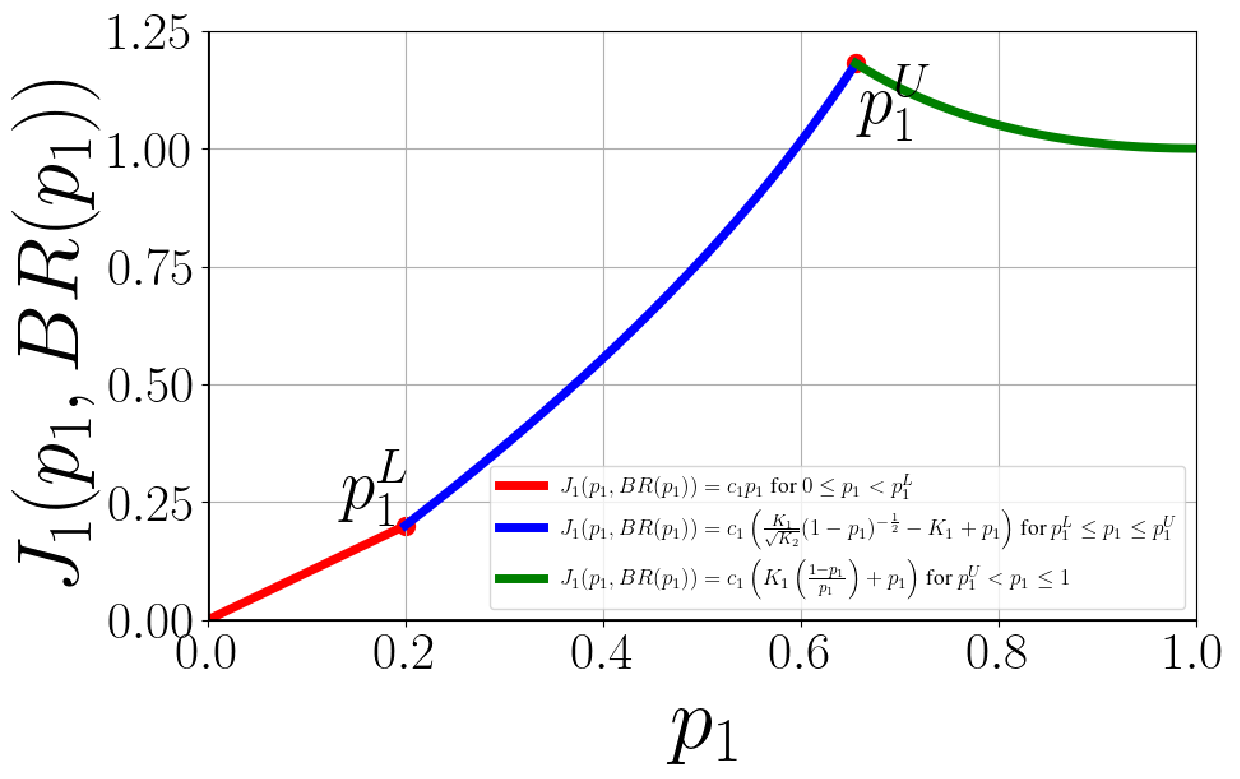}}
		\subfigure[]{%
			\includegraphics[scale=0.17]{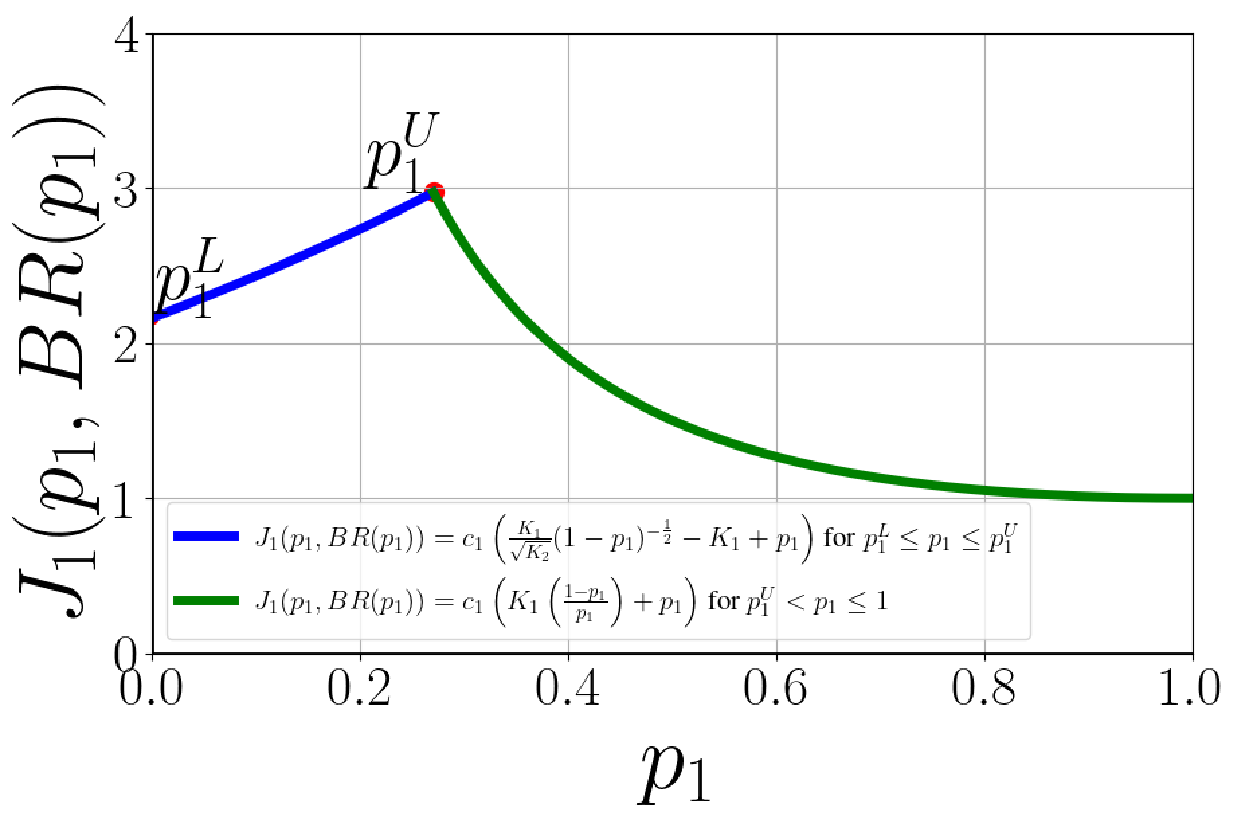}}\\ \vspace{-0.35cm}
   \subfigure[]{%
			\includegraphics[scale=0.17]{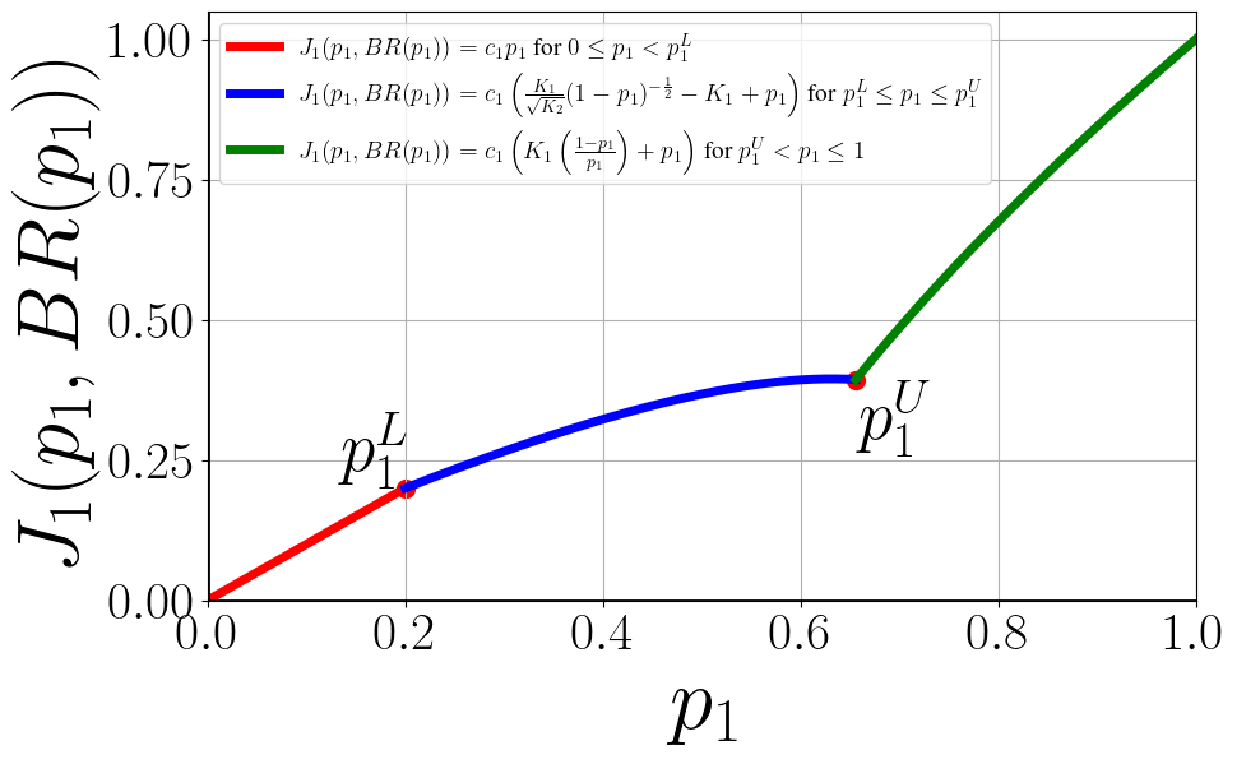}}
   \subfigure[]{%
			\includegraphics[scale=0.17]{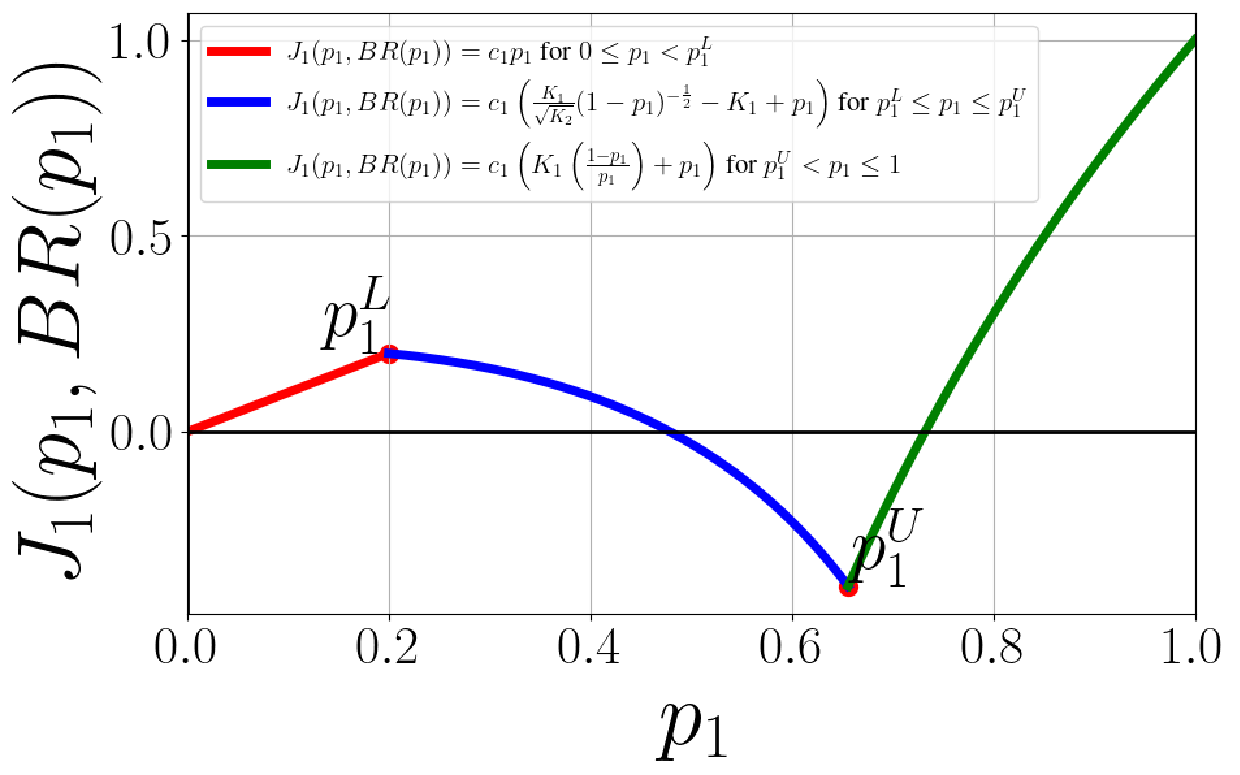}}
	\end{center}\vspace{-0.5cm}
	\caption{The plots of the objective function $J_1(p_1, BR(p_1))$ for (a) $K_2=1.25$ and $K_1>0$, (b) $K_2=0.1$ and $K_1>0$, (c)$K_2>0$ and $K_1=-0.5$, and (d) $K_2>0$ and $ K_1 = -2$. }
	\label{fig:sim1-sec}
	\vspace{-0.3cm}
\end{figure}





In this section, we provide numerical results to simulate the behavior of $J_1(p_1, BR(p_1))$ in (\ref{eqn:cost_player_1_K_2>0}) when $K_2>0$. In all simulation results, we take $c_1=1$. In the first numerical simulations, we consider $K_2=1.25$ and $K_1=1$ and plot $J_1(p_1, BR(p_1))$ in Fig.~\ref{fig:sim1-sec}(a). As stated in Corollary~\ref{Cor_8}, when $K_2>1$ and $K_1>0$, $J_1(p_1, BR(p_1))$ is increasing in $0\leq p_1\leq p_1^U$, and $J_1(p_1, BR(p_1))>0$ when $ p_1^U<p_1\leq 1$, thus the SE is attained at $(p_1^*,BR(p_1^*)) =(0,1)$ and the minimum cost of player $P_1$ is  $J_1(0,1)=0$. For the second numerical experiments, we consider $K_2=0.1$ and $K_1=1$. This is the case where $K_2=0.1$ which could be due to the higher sampling cost $c_2$, disincentivizing player $P_2$ from sampling. We plot $J_1(p_1, BR(p_1))$ in Fig.~\ref{fig:sim1-sec}(b). We see that the SE happens $(p_1^*,BR(p_1^*)) =(1,0)$ as this policy lowers the cost compared to free-riding policy which was the SE in the previous case. In the last 2 simulation results, we consider $K_2=1.25$ for both simulations, but take $K_1=-0.5$ and $K_1=-2$, respectively. When $K_2>0$ and $K_1<0$, we may have two different equilibrium policies. For $K_1=-0.5$, shown in Fig.~\ref{fig:sim1-sec}(c), the SE policy is obtained at $p_1^*=0$. On the other hand, for $K_1=2$, shown in Fig.~\ref{fig:sim1-sec}(d), the SE policy is obtained at $p_1^*=p_1^U$.


\vspace{-0.35cm}
\section{Discussion and Conclusion}\label{Sec:discussion}
\vspace{-0.4cm}
To summarize, when $K_1$ and $K_2$ are positive, both players care more about estimating their opponent's state. Due to the commitment power, the leader, player $P_1$, can \textit{generally} force the follower, player $P_2$, to sample, while staying silent, which is $p_1^*=0$, $p_2^*>0$ and is observed in Fig.~\ref{fig:sim1-sec}(a). However, for some specific cases where $K_2$ is relatively small compared to $K_1$, which implies that the cost of sampling for $P_1$ is relatively low, free-riding for the leader is more costly than sampling. The relative disadvantage of sampling causes player $P_2$ not to sample as the best response. On the other hand, the considerably lower sampling cost inclines the leader to sample. Thus, we may have  $p_1^*>0$, $p_2^*=0$ as observed in Fig.~\ref{fig:sim1-sec}(b).

When $K_1$ is negative, but $K_2$ is positive, an intuitive strategy for player $P_1$ is not to sample, since the importance of concealing information is higher, which may imply $p_1^*=0$, $p_2^*>0$ is the SE (which is in fact the case in Fig.~\ref{fig:sim1-sec}(c)). However, player $P_1$ might still sample when the magnitude of $K_1$ is high compared to $K_2$. The cost of sampling of player $P_1$ ($c_1$) is inversely proportional to $K_1$. Thus, when $c_1$ is considerably low, it is more profitable for the leader to sample instead of free-riding. Thus, $p_1^*>0$, $p_2^*=0$ as shown in Fig.~\ref{fig:sim1-sec}(d). When $K_1$ is positive, but $K_2$ is negative, player $P_2$ does not sample because the follower cares more about revelation of the information. The leader samples with $\min\{\sqrt{K_1}, 1\}$, where the SE is given by $p_1^*>0$, $p_2^*=0$. However, as the cost of sampling of player $P_1$ increases, $\sqrt{K_1}$ decreases, i.e.,  the sampling probability of the leader gets closer to zero. 

In this paper, we have studied remote estimation games among 2 players, interested in estimating the other player's state as timely as possible while avoiding revealing of self information to the other player and the cost of updating. Here, the players can make an observation about their opponent's state; however, it also reveals information about the player's own state, creating an intricacy between sampling vs. revealing information. By focusing our attention on stationary probabilistic sampling policies, we have obtained the characterization of the SE of the game.  
\vspace{-0.3cm}
\bibliography{references}
\end{document}